\documentclass[reprint,aps,prl,footinbib,superscriptaddress,floatfix,twocolumn]{revtex4-2}
\usepackage{graphics}
\usepackage{graphicx}
\usepackage{subfigure}
\usepackage{mathrsfs}
\usepackage{amsmath}
\usepackage{color}
\usepackage{natbib}
\usepackage{bookmark}

\begin{document}
\date{\today}

\title{Multiflagellate Swimming Controlled by Interflagella Hydrodynamic Interactions}

\author{Shiyuan Hu}
\email[]{shiyuan.hu@itp.ac.cn}
\affiliation{CAS Key Laboratory of Theoretical Physics, Institute of Theoretical Physics, Chinese Academy of Sciences, Beijing 100190, China}
\author{Fanlong Meng}
\email[]{fanlong.meng@itp.ac.cn}
\affiliation{CAS Key Laboratory of Theoretical Physics, Institute of Theoretical Physics, Chinese Academy of Sciences, Beijing 100190, China}
\affiliation{School of Physical Sciences, University of Chinese Academy of Sciences, 19A Yuquan Road, Beijing 100049, China}
\affiliation{Wenzhou Institute, University of Chinese Academy of Sciences, Wenzhou, Zhejiang 325000, China}

\begin{abstract}
Many eukaryotic microorganisms propelled by multiple flagella can swim very rapidly with distinct gaits. Here, we model a three-dimensional mutiflagellate swimming strategy, resembling the microalgae, and investigate the effects of interflagella hydrodynamic interactions (iHIs) on the swimming performance. When the flagella are actuated synchronously, the swimming efficiency can be enhanced or reduced by iHIs, determined by the intrinsic tilting angle of the flagella. The asynchronous gait with a phase difference between neighboring flagella is found to be important by both utilizing the iHIs and reducing the oscillatory motion via the basal mechanical coupling. We further demonstrate that an optimal number of flagella could arise when the microswimmer is loaded with a swimmer body. Apart from understanding the role of iHIs in the multiflagellate swimming, this work could also guide laboratory fabrications of novel microswimmers.
\end{abstract}
\maketitle

Swimmers propelled by multiple slender appendages are ubiquitous in nature, from microorganisms~\cite{Lauga09, Elgeti15} to macroscopic ones including crustaceans and feather stars~\cite{[{See a video showing a swimming feather star with a fascinating asynchronous beating pattern. }] Petri16, Madin22}. These examples trigger numerous research interests in the emergence of collective motions, mediated via long-range hydrodynamic interactions~\cite{Uchida10, Guo18, Han18, Chakrabarti19, Man20, Meng21, Chakrabarti22, Kanale22, Tuatulea22, Hickey23}, short-range steric interactions~\cite{Chelakkot21}, and the mechanical couplings through cell body motion~\cite{Friedrich12, Geyer13} and intracellular fibers~\cite{Quaranta15, Wan16, Liu18, Guo21}. Meanwhile, searching the optimal gaits for swimming helps the understanding of biological locomotion mechanisms~\cite{Tam11, Osterman11, Eloy12} and inspires diverse designs of artificial microswimmers~\cite{Dreyfus05, Ye13, Lum16, Lu18, Diaz21, Wang23}.

\begin{figure}[b]
\centering
\includegraphics[bb= 0 16 365 140, scale=0.67,draft=false]{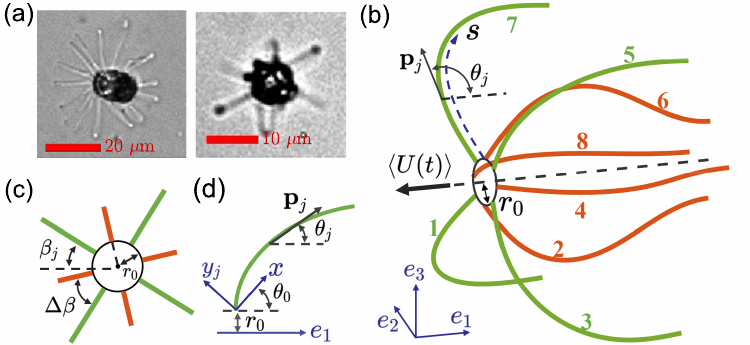}
\caption{(a) Swimming hexadecaflagellate \textit{P. cyrtoptera} with a synchronous gait (front view, left panel), and octoflagellate \textit{P. octopus} with an asynchronous gait (side view, right panel), adapted from~\cite{Wan16}. (b) A multiflagellate microswimmer model. The lab frame is $\{e_1, e_2, e_3\}$. (c) Front view of the microswimmer. (d) The filament frame $\{x, y_j\}$ lies in each beating plane. The filament oscillates around $\hat{\textbf{x}}$. For $\theta_0=0$, $\hat{\textbf{x}}$ aligns with $\hat{\textbf{e}}_1$.}
\label{fig1}
\end{figure}

At low Reynolds number (Re), due to the long-range decay of flow velocities, the motions of the biological filaments are strongly affected by the flows created by other {filaments} nearby. Compared to prokaryotic cells in which multiple flagella usually bundle together while rotating~\cite{Berg04}, the eukaryotic flagella perform well-separated beating motions. Until now, the studies about the effects of eukaryotic interflagella hydrodynamic interactions (iHIs) on swimming have been limited to two coplanar flagella~\cite{Singh18, Elfasi18, Liu20}. However, eukaryotic cells can also grow multiple flagella, exemplified by the unicellular algae, for instance, species \textit{Polytomella} and \textit{Pyramimomas}~\cite{De81, Hori87, Daugbjerg92, Wan16}. These algal flagella are anchored at the anterior pit of the cell body beating in different crossing planes [Fig.~\ref{fig1}(a)]. To further complicate the iHIs, the flagella can beat with distinct gaits~\cite{Wan16, Wan18, Wan20}, which include a synchronous gait with all flagella in phase and an asynchronous gait with neighboring filaments out of phase. Experiments have shown that the asynchronous gait generates a faster swimming than other gaits~\cite{Wan16, Diaz21}, but how iHIs determine the swimming performance is still unclear. 

In this letter, we study the effects of the interflagella hydrodynamic interactions in a multiflagellate microswimmer at low Re, combining numerical simulations and a small-amplitude analytical theory. We find that the iHIs are controlled by the intrinsic flagella orientation and the phase difference between neighboring flagella, which in turn determine the swimming performance and the advantage of propelling with multiple flagella.

\emph{Microswimmer modelling}.---
We model the flagella as slender filaments of radius $a$, length $L$ ($a/L \ll 1$), and bending rigidity $B$, moving in a Stokesian fluid of viscosity $\mu$. The filaments are anchored on a frictionless circle of radius $r_0$ [Fig.~\ref{fig1}(b)]. Here, we neglect the small nonplanar component of the flagella beating for simplicity, which are usually responsible for the slow rotation of algae during swimming~\cite{Cortese21}, and constrain the filaments in evenly spaced two-dimensional planes [Fig.~\ref{fig1}(c)]. To ensure a straight swimming trajectory, the opposed filaments are always kept in the same phase, i.e., mirror-symmetric about the central axis, but neighboring filaments can be actuated with different phases. The supporting circle can transmit forces and constrain one ends of the filaments to move at the same velocity, resembling the mechanical coupling of flagella via the cell body~\cite{Friedrich12}. Motivated by the bending-wave deformation of natural flagella, we adopt a simple actuation mechanism by oscillating the filament tangential angles~\cite{Yu06, Lauga07, Peng17} at the anchored points with amplitude $\theta_{\mathrm{A}}$ around an intrinsic tilting angle $\theta_0$, defined relatively to the swimming direction [Fig.~\ref{fig1}(d)].

\footnotetext[1]{See Supplemental Material for simulation videos, numerical and analytical details, which includes Refs.~\cite{Hori87, Wiggins98_2,Lauga07,Wan18,Hu23,Quaranta15,Xu16}.}

The filaments are mathematically modelled as Euler-Bernoulli beams with force density $\textbf{f}_j = -B\partial_s^4 \textbf{r}_j + \partial_s(T\textbf{p}_j)$, where $\textbf{r}_j(s,t)$ is centerline position with $s$ the arc length, $\textbf{p}_j$ is the tangent vector, and $T$ is the filament tension. From a nonlocal slender-body theory, the dynamics of the filaments are governed by a set of coupled equations~\cite{Johnson80, Tornberg04},
\begin{equation}\label{eq1}
8\pi\mu (\partial\textbf{r}_j/\partial t - \textbf{v}_j) = \Lambda[\textbf{f}_j] + K[\textbf{f}_j],
\end{equation}
where the background flow $\textbf{v}_j$ felt by filament $j$ includes the nonlocal Stokeslet flows generated by all other filaments, and $\Lambda[\textbf{f}_j]$ and $K[\textbf{f}_j]$ capture the local and nonlocal effects within each filament, respectively. The tangential angles at the anchored points are actuated periodically as $\theta_j(s=0, t) = \theta_{\mathrm{A}}\sin (2\pi t/\tau +\phi_j) + \theta_0$ with $\tau$ the period and $\phi_j$ the initial phase. The relative importance of viscous to elastic stresses on the filament is characterized by the elastoviscous number $\eta = L/(B\tau/8\pi\mu)^{1/4}$; by considering physical parameters of realistic flagella, we take $\eta=3$ for most simulations. Equation~(\ref{eq1}) is solved numerically, subjected to the mechanical coupling at $s=0$ and the zero-force and -torque constraints at $s=L$ (see Sec. S2 in Supplemental Material~\cite{Note1} for details).

\begin{figure}[t]
\centering
\includegraphics[bb= 0 15 360 428, scale=0.67,draft=false]{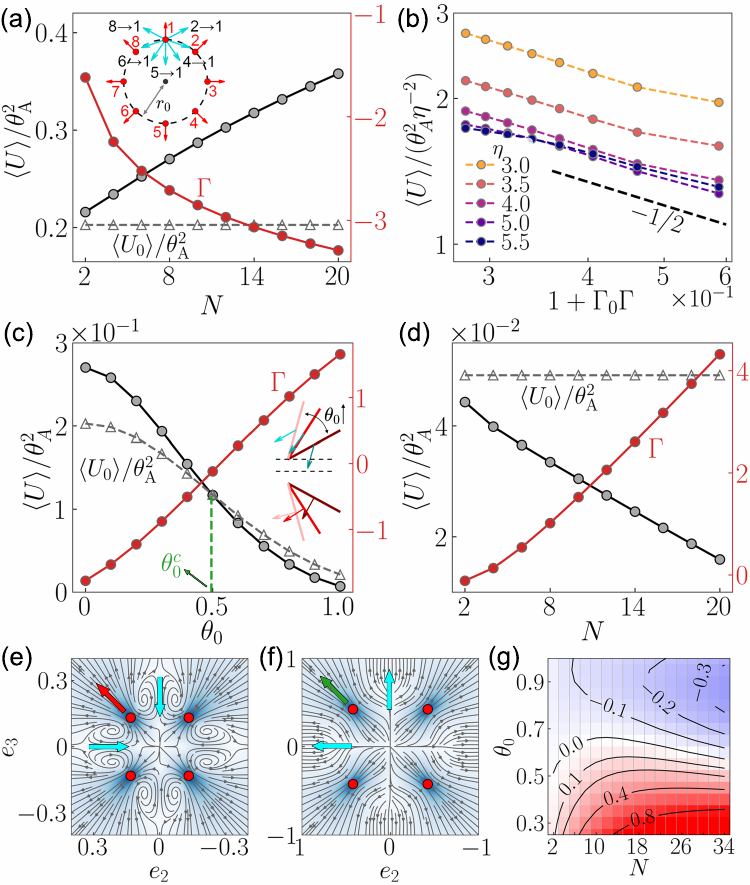}
\caption{Effects of iHIs for the synchronous gait with $r_0/L = 0.1$--$0.2$ ($\Gamma_0\approx 0.2$--$0.3$). (a) $\langle U \rangle/\theta_{\mathrm{A}}^2$ (dark circles, left axis), $\langle U_0 \rangle/\theta_{\mathrm{A}}^2$ (open triangles), and $\Gamma$ (red circles, right axis) versus $N$ for $\theta_0 = 0$ and $\theta_{\mathrm{A}}=0.2$. (b) $\langle U\rangle/\theta_{\mathrm{A}}^2\eta^{-2}$ versus $1+\Gamma_0\Gamma_y$ in log-log scale. (c), (d) $\langle U\rangle/\theta_{\mathrm{A}}^2$, $\langle U_0\rangle/\theta_{\mathrm{A}}^2$, and $\Gamma$ (c) versus $\theta_0$ with fixed $N = 8$ and (d) versus $N$ with fixed $\theta_0 = 0.8$. (e), (f) Cross-sectional velocity field for (e) $\theta_0 = 0$ and (j) $\theta_0 = 0.8$. The red dots are the positions where the cross section intersects with the filaments. (g) $(\mathcal{E}-\mathcal{E}_0)/\mathcal{E}_0$ versus $\theta_0$ and $N$ for $\theta_{\mathrm{A}} = \theta_0$. Insets in (a) and (c) show the schematics of $\Gamma_{jk}$ for (a) $\theta_0 = 0$ and (c) $\theta_0 > 0$. The red arrows indicate the filament forces $\textbf{f}_k$, represented as point forces, and the cyan arrows indicate $\textbf{v}_{k \to j}$.}
\label{fig2}
\end{figure}
\emph{Synchronous beating}.---In the synchronous gait, all filaments have the identical planar deformations and bending forces. From numerical simulations based on Eqs.~(\ref{eq1}), we observe that the instantaneous swimming speed $U(t)$, measured as the speed of the supporting circle, oscillates with time, but on time average the swimmer translates with a net speed $\langle U(t)\rangle$ along the $-e_1$ direction. 

To analyze the swimming behavior, we first consider the case of $\theta_0 = 0$. As shown in Fig.~\ref{fig2}(a), $\langle U\rangle$ increases with the filament number $N$; but the swimming speed without incorporating iHIs $\langle U_0\rangle$ ($\textbf{v}_j = 0$) does not change with $N$, since the total propulsion increases in proportion to the total viscous drag as $N$ increases. In our model, the swimming dynamics is influenced by the iHIs in two ways: the drift flow along the swimming direction $v_{e_1}(t) = \textbf{v}_j(s=0, t)\cdot \hat{\textbf{e}}_1$ contributes directly to the instantaneous $U(t)$, and the relative motion between the filament beating and the flow perpendicular to the filament changes the filament deformations and forces, which generates propulsion relatively to the background flow. One can easily identify that for small tilting angles $\theta_0 \ll 1$, $v_{e_1}$ is small due to the small filament forces along $\hat{\textbf{e}}_1$. Therefore, the key to understand the effect of iHIs lies with the perpendicular flow. In order to characterize this effect, we consider the small-amplitude limit $\theta_{\mathrm{A}} \ll 1$. Denote the frame translating with the filament as $\{x, y_j\}$ [Fig.~\ref{fig1}(d)]. At the leading order in $\theta_{\mathrm{A}}$, $s\approx x$, and the planar deformations $y_j(x,t)$ satisfy the linear equations~\cite{Wiggins98_1, Wiggins98_2}, $\xi_{\perp}\left(\partial y_j/\partial t - v_j^{\perp}\right) - f_j = 0$, where the perpendicular drag coefficient $\xi_{\perp} \approx 4\pi\mu/\ln(L/a)$, $v_j^{\perp} = \textbf{v}_j \cdot \hat{\textbf{y}}_j$, and the filament bending force $f_j(x,t) = -B\partial^4 y_j/\partial x^4$. 

To fully describe $y_j(x,t)$, we further approximate the nonlocal Stokeslet flow using a local velocity-force relation: for the flow velocity at filament $j$ generated by filament $k$,
\begin{equation}\label{eq2}
v_{k\to j}^{\perp}(x,t) = \textbf{v}_{k\to j}\cdot \hat{\textbf{y}}_j \approx \Gamma_{jk} \ln(L/r_0) f_k(x,t)/(4\pi\mu),
\end{equation}
where $\Gamma_{jk}$ denotes the coupling strength. Equation~(\ref{eq2}) is valid for $r_0/L \ll 1$~\cite{Goldstein16, Man16}. Different from the previous coplanar case, $\Gamma_{jk}$ accounts for the relative orientation between the beating planes of filaments $j$ and $k$. By denoting  $\lambda_{jk} = \cos(\beta_j - \beta_k)$ with $\beta_j$ the angle between the beating plane and the $e_1$-$e_2$ plane [Fig.~\ref{fig1}(c)], the coupling strength can be derived as (Sec. S3 in~\cite{Note1})
\begin{equation}\label{eq3}
\Gamma_{jk} = \frac{1}{2\ln (r_0/L)}\left[1+\lambda_{jk}\ln \left(\frac{1-\lambda_{jk}}{2e} \frac{r_0^2}{L^2}\right)\right],
\end{equation}
which essentially quantifies the alignment of $v^{\perp}_{k\to j}\hat{\textbf{y}}_j$ with the motion of filament $j$ itself: they are opposite for $\Gamma_{jk} < 0$ but aligned with each other for $\Gamma_{jk} > 0$. Then the total velocity $v^{\perp}_j = \sum_k v^{\perp}_{k \to j} = \Gamma \ln (L/r_0)f_j/(4\pi\mu)$, where the total coupling strength $\Gamma = \sum_{k \ne j}^N \Gamma_{jk}< 0$. A schematic of $\Gamma$ is shown in Fig.~\ref{fig2}(a) inset for $\theta_0 = 0$ and $N = 8$, where $\Gamma$ is negative. 

By substituting $v^{\perp}_j$, we can solve for $y_j(x,t)$; with the solution, we deduce that at large $\eta$ (Sec. S4 in~\cite{Note1})
\begin{equation}\label{eq4}
\langle U \rangle \sim (L/\tau)\theta_{\mathrm{A}}^2\eta^{-2} (1+\Gamma_0\Gamma)^{-1/2},
\end{equation}
where we define the intrinsic coupling strength $\Gamma_0 = \ln (L/r_0)/\ln (L/a)$. From this simple relation, $\langle U \rangle$ increases as $\Gamma$ decreases, which is verified by numerical simulations [Fig.~\ref{fig2}(a)]; the $-1/2$ exponent is also confirmed in Fig.~\ref{fig2}(b). For $\Gamma<0$, the filaments are moving against an opposing flow generated by all other filaments during both the recovery and power strokes and are bent more than those in a quiescent fluid. Consequently, the projection of the bending forces onto the swimming direction leads to an enhanced propulsion and swimming speed.

To characterize the swimming efficiency, we introduce $\mathcal{E} = D_{\mathrm{f}}\langle U\rangle^2 /\langle P\rangle$~\cite{Lighthill52, Becker03}, where $D_{\mathrm{f}}$ is resistivity of a translating filament with fixed $\theta(s) = \theta_0$ and the average power against the viscous fluid is $\langle P\rangle = \langle\int_0^L\partial\textbf{r}_j/\partial t\cdot \textbf{f}_j\,ds \rangle$. For $\theta_{\mathrm{A}} \ll 1$ and $\theta_0 = 0$, $\langle P\rangle = \langle \int_0^L f_j \partial y_j/\partial t\,dx \rangle$; at large $\eta$, $\langle P\rangle \sim \theta_{\mathrm{A}}^2 \eta B(\tau L)^{-1} (1+\Gamma_0\Gamma)^{-1/4}$, i.e., the viscous dissipation is increased by iHIs for $\Gamma < 0$ and decreased for $\Gamma>0$. Combining the scaling of $\langle U \rangle$, we can obtain $\mathcal{E}\sim \theta_{\mathrm{A}}^2\eta^{-1}(1+\Gamma_0 \Gamma)^{-3/4}$, and therefore $\mathcal{E}$ is also enhanced when $\Gamma$ is negative.

For cases of $\theta_0 > 0$, $\langle U\rangle$ decreases with increasing $\theta_0$ and becomes smaller than $\langle U_0\rangle$ when $\theta_0$ is larger than a critical value $\theta_0^{c} \simeq 0.5$ [Fig.~\ref{fig2}(c)]. This result coincides with the geometrical dependence of the coupling strength. Since the coupling strength between neighboring filaments are always positive, the total coupling strength $\Gamma$ is mainly affected by the opposed filaments. As depicted in Fig.~\ref{fig2}(c) inset, for an opposed pair $j$ and $k$, $\textbf{v}_{k \to j}$ induced by $f_k \hat{\textbf{y}}_k$ is more aligned with the orientation of filament $j$ as $\theta_0$ increases, and at sufficiently large $\theta_0$, $v_{k \to j}$ turns positive. To estimate $\Gamma_{jk}$ in this case, we approximate $f_k$ as a constant force density of strength $B/L^3$ and compute $v_{k \to j}$ numerically. Then the coupling strength is calculated as
\begin{equation}\label{eq5}
\Gamma_{jk} = \frac{1}{2\ln(L/r_0)}\frac{\eta^4\tau}{L^2}\int_0^L v_{k \to j}\,ds,
\end{equation}
which increases and turns positive as $\theta_0$ increases [Fig.~\ref{fig2}(c)]. Such effects can also be observed from the flow field in the $e_2$-$e_3$ plane, which points outward from the central axis during the recovery stroke [Fig.~\ref{fig2}(f)]. This is in contrast to the case of $\theta_0 = 0$, where the flow field between neighboring filaments points inward [Fig.~\ref{fig2}(e)]. Meanwhile, for fixed $\theta_0 > \theta_0^{c}$, e.g., 0.8, $\langle U\rangle$ decreases with the number of the filaments $N$, accompanied with an increase in $\Gamma$ [Fig.~\ref{fig2}(d)]. Figure~\ref{fig2}(g) shows the change of swimming efficiency by iHIs, $(\mathcal{E}-\mathcal{E}_0)/\mathcal{E}_0$, across a relatively wide range of $\theta_0$ and $N$. In the above analysis, we have neglected the drift flow in the swimming direction $v_{e_1}(t)$, which can be important for large $\theta_0$ and $r_0/L\rightarrow 0$. But in this work, we focus on the case of $r_0/L \gtrsim 0.1$, where the effect of iHIs is mainly encapsulated in $\Gamma$ (Sec. S5 in~\cite{Note1}).

\begin{figure}[t]
\centering
\includegraphics[bb= 0 10 365 308, scale=0.67,draft=false]{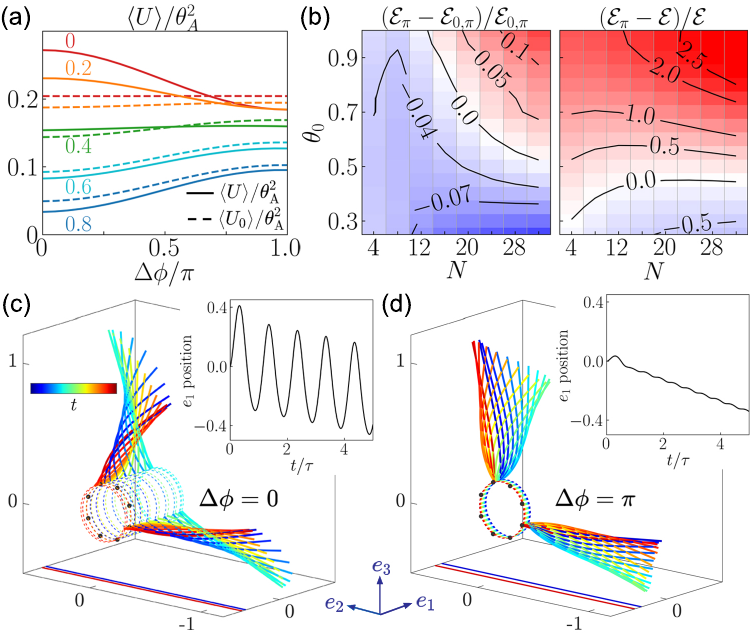}
\caption{Effect of asynchronous gait on the swimming performance. (a) $\langle U \rangle/\theta_{\mathrm{A}}^2$ and $\langle U_0 \rangle/\theta_{\mathrm{A}}^2$ versus $\Delta\phi$ for different values of $\theta_0$ ($\theta_0 = 0$, 0.2, 0.4, 0.6, and 0.8) with fixed $N = 8$ and $\theta_{\mathrm{A}} = 0.1$. (b) $(\mathcal{E}_{\pi}-\mathcal{E}_{0,\pi})/\mathcal{E}_{0,\pi}$ and (c) $(\mathcal{E}_{\pi}-\mathcal{E})/\mathcal{E}$ as a function of $\theta_0$ and $N$ for $\theta_{\mathrm{A}} = \theta_0$. (d), (e) Time lapse of a swimmer with $N=8$ over an actuation period for (d) $\Delta\phi=0$ and (e) $\Delta\phi=\pi$ with $\theta_{\mathrm{A}} = \theta_0 = 1.0$. Time runs from blue to red. Due to symmetry only two filaments are shown. The dark dots are the anchored points. Insets show the $e_1$ position of the supporting circle for five actuation periods. The blue and red lines mark the initial and final $e_1$ positions. See Sec. S1 in~\cite{Note1} for simulation videos.}
\label{fig3}
\end{figure}

\emph{Asynchronous gait}.---
In the asynchronous gait, we set $\phi_j = \phi_{j+2}$, and neighboring filaments beat with a phase difference, $\Delta\phi = |\phi_j - \phi_{j+1}| \ne 0$. The simulation results show that the swimming speed incorporating iHIs $\langle U\rangle$ is strongly affected by $\Delta\phi$. As shown in Fig.~\ref{fig3}(a), for large $\theta_0 = 0.6$ and $0.8$, $\langle U\rangle$ increases with $\Delta\phi$; for small $\theta_0 = 0$ and $0.2$, $\langle U \rangle$ decreases with $\Delta\phi$; for intermediate $\theta_0 = 0.4$, $\langle U \rangle$ remains nearly unchanged. 

The first mechanism responsible for the $\Delta\phi$-dependence of the swimming speed is the mechanical coupling arising from the force-free condition and the rigid constraint imposed by the supporting circle along the $e_1$ direction, which is absent in the synchronous case and independent from iHIs. Compared to the synchronous case, the oscillatory motion of the swimmer is significantly suppressed by the mechanical coupling [Figs.~\ref{fig3}(c) and ~\ref{fig3}(d)]. Physically, this is reminiscent of Huygens' double pendula~\cite{Pikovsky01}, with the difference that in our model the phase difference is fixed rather than evolving through a synchronization process. To quantify this effect, we consider a simple model of rigid filaments ($\eta \to 0$) without iHIs, which can not achieve net motion and only oscillates given imposed angular dynamics $\theta_j (t)$. Integration of Eqs.~(\ref{eq1}) relates the total filament force $\textbf{F}_j$ with $U(t)$ and $\theta_j(t)$. By enforcing $\sum_j \textbf{F}_j \cdot \hat{\textbf{e}}_1 = 0$, we obtain the oscillation amplitude of the swimming speed $\Delta U$; in the leading order of $\theta_{\mathrm{A}}$,
\begin{equation}\label{eq6}
\Delta U = \pi\theta_{\mathrm{A}} \frac{L}{\tau} \frac{\gamma \sin\theta_0}{\cos^2\theta_0+\gamma\sin^2\theta_0}\cos(\Delta\phi/2),
\end{equation}
which decreases to zero as $\Delta\phi$ increases from 0 to $\pi$. Equation~(\ref{eq6}) also points out that $\Delta U = 0$ for $\theta_0 = 0$, which indicates that the effect of mechanical coupling is small for small $\theta_0$. With reduced oscillations, the swimming speed without incorporating iHIs $\langle U_0 \rangle$ for $\eta = 3$ increases with $\Delta\phi$ for $\theta_0 > 0$, as shown in Fig.~\ref{fig3}(a). 

By comparing $\langle U \rangle$ and $\langle U_0 \rangle$, it is now evident that the enhanced swimming speed at large $\theta_0$ by the asynchronous gait is mainly due to the mechanical coupling. However, it can not explain the decreasing of $\langle U \rangle$ observed at small $\theta_0$, where iHIs dominate. We next exploit the concept of coupling strength to present an intuitive understanding. We have already seen in the synchronous case that $\langle U \rangle$ is enhanced by iHIs with $\Gamma$ negative at small $\theta_0$ [Fig.~\ref{fig2}]. Now consider the case of $\Delta\phi = \pi$. This antiphase beating will turn part of the total coupling strength $\Gamma$ from negative to positive by reversing the beating directions of half of the filaments; since $\langle U \rangle$ is inversely related with $\Gamma$ [Eq.~(\ref{eq4})], it will be decreased. An analytical calculation can be performed based on the small-$\theta_{\mathrm{A}}$ approximation to quantitatively account for the effect of the full range of $\Delta\phi \in [0, \pi]$ (Sec. S4 in~\cite{Note1}). 

By further comparing the swimming efficiency of the antiphase case with iHIs $\mathcal{E}_{\pi}$ and that without iHIs $\mathcal{E}_{0,\pi}$, one can see that iHIs enhance the swimming efficiency at large $\theta_0$ and large $N$ [Fig.~\ref{fig3}(b) left panel], which is roughly in reverse to the synchronous case shown in Fig.~\ref{fig2}(g). Meanwhile, we quantify the combined effect of iHIs and mechanical coupling by comparing $\mathcal{E}_{\pi}$ and efficiency of the synchronous case with iHIs $\mathcal{E}$, as shown in Fig.~\ref{fig3}(b) right panel. For small $\theta_0$, $\mathcal{E}_{\pi} < \mathcal{E}$ due to iHIs; but for the more practical case of large $\theta_0$, $\mathcal{E}_{\pi}$ is significantly larger than $\mathcal{E}$, and the contribution of the mechanical coupling to the enhanced efficiency dominates over that of iHIs. 

\footnotetext[2]{The cell body is elongated with aspect ratio $b/l \approx 0.5$, where the polar radius $l \sim L$~\cite{Wan18}. The drag coefficient of a spheroid is $6\pi\mu b H(b/l)$, where the factor $H(b/l) \approx 1.2$ for $b/l \approx 0.5$~\cite{Happel12}. Therefore, the effective radius $b H(b/l) \approx 0.6 L$.}

\begin{figure}[t]
\centering
\includegraphics[bb= 0 15 360 158, scale=0.67, draft=false]{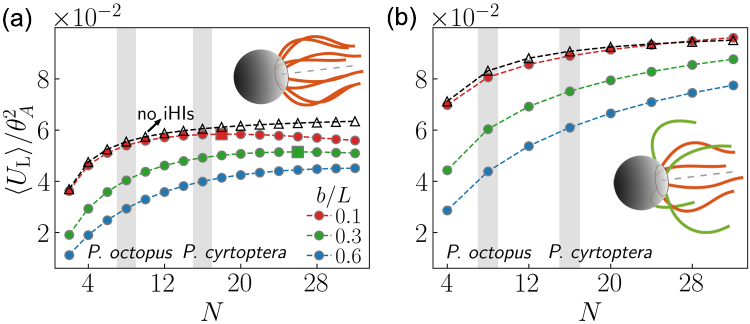}
\caption{Time-averaged swimming speed of a loaded microswimmer $\langle U_{\mathrm{L}} \rangle/\theta_{\mathrm{A}}^2$ as a function of the filament number $N$ for (a) synchronous gait and (b) asynchronous gait with $\Delta\phi=\pi$. Open triangles show results without iHIs for $b/L= 0.1$. Squares in (a) indicate the optimal filament numbers. Other simulation parameters are $\theta_0 = \theta_{\mathrm{A}} = 0.8$.}
\label{fig4}
\end{figure}

\emph{Loaded microswimmers}.---We examine the swimming performance of the microswimmer when loaded with a passive sphere. The filaments are anchored on the posterior of a sphere of radius $b$ [Fig.~\ref{fig4} insets]. Balancing the propulsion generated without the load with the combined viscous drag of the filaments and the load, we obtain the time-averaged speed, $\langle U_{\mathrm{L}} \rangle = D_{\mathrm{f}}\langle U \rangle / (D_{\mathrm{f}} + 6\pi\mu b/N)$, which is valid at the order of $\theta_{\mathrm{A}}^2$ by ignoring the hydrodynamic interactions between the sphere and the filaments (Sec. S4 in~\cite{Note1}). In the synchronous gait where iHIs impede the swimming of the non-loaded swimmer [Fig.~\ref{fig2}], $\langle U_{\mathrm{L}}\rangle$ has a maximum at an optimal filament number [Fig.~\ref{fig4}(a)]. This is attributed to the competition between the reduced load shared by each filament as $N$ increases and the decreased propulsion per filament due to iHIs. The optimal filament number increases with the increase of the sphere size. By contrast, for the asynchronous gait $\langle U_{\mathrm{L}} \rangle$ increases monotonically, since the effect of mechanical coupling dominates [Fig.~\ref{fig3}(b)]. 

For algal cells with effective body radius $b/L \approx 0.6$~\cite{Note2} and filament number $N=2$--16, Fig.~\ref{fig4} indicates that their swimming speed keeps increasing as $N$ increases without attaining a maximum for both the synchronous and asynchronous gaits. Note that the above analysis is only qualitative; to obtain quantitative conclusions, the interaction between the no-slip surface of the swimmer body and the filament dynamics needs to be considered~\cite{Liu20, Wrobel16}. Perhaps the most important effect is that if the filaments are anchored on the anterior of the body, the no-slip surface will partially screen the iHIs, especially between the opposed filaments.

In conclusion, we demonstrate the effect of interflagella hydrodynamic interactions on the swimming performance of a multiflagellate swimmer. We find that iHIs can either promote or impede the swimming performance, depending on the filament tilting angle and the phase difference between neighboring filaments. Our results have implications for the motility of microalgae and suggest novel designs of artificial microswimmers using soft materials~\cite{Zhou21}. 

An experiment of a rotating microswimmer also finds that sufficient filaments can impede swimming~\cite{Ye13}; although the filaments are revolving around the central axis rather than beating, both opposed and aligned filament motions are present. The physical picture based on the hydrodynamic coupling strength developed in this work may also be relevant there. The bacteria swimming speed also varies nonlinearly with the flagella number~\cite{Nguyen18}, but the short-range viscous and steric interactions are very important within the flagella bundle~\cite{Kamdar22}. The microswimmer may benefit from more phase-shifted groups, as seen in \textit{P. octopus}~\cite{Wan16}. Indeed, asymptotic calculations without iHIs have shown that three pairs of breaststrokes with different phases is sufficient to generate a steady swimming speed~\cite{Lauga07}. Finally, the waveform of the common breaststroke of algal flagella has a static component, which can be approximated as a circular arc with a constant curvature~\cite{Geyer16}. Our limited simulations with a nonzero intrinsic curvature show that the swimming performance is improved significantly~\cite{Note1}, which suggests a possible approach to tune the iHIs~\cite{Liu20, Liu20_2, Hu22}.

\bigskip

\begin{acknowledgments}
We thank Michael Shelley, Jun Zhang, and Masao Doi for helpful conversations. We thank Hugues Chat\'e for suggestions on the manuscript. This work is supported by National Natural Science Foundation of China (Grant No. 12275332, 12047503, and 12247130), Chinese Academy of Sciences, Max Planck Society and Wenzhou Institute (Grant No. WIUCASQD2023009).
\end{acknowledgments}

\bibliography{reference}

\end{document}


\date{\today}
\title{Supplemental Material for \\
Multiflagellate Swimming Controlled by Interflagella Hydrodynamic Interactions}

\author{Shiyuan Hu}
\affiliation{CAS Key Laboratory of Theoretical Physics, Institute of Theoretical Physics, Chinese Academy of Sciences, Beijing 100190, China}
\author{Fanlong Meng}
\affiliation{CAS Key Laboratory of Theoretical Physics, Institute of Theoretical Physics, Chinese Academy of Sciences, Beijing 100190, China}
\affiliation{School of Physical Sciences, University of Chinese Academy of Sciences, 19A Yuquan Road, Beijing 100049, China}
\affiliation{Wenzhou Institute, University of Chinese Academy of Sciences, Wenzhou, Zhejiang 325000, China}

\maketitle

\section{Supplemental videos}
Videos 1 and 2: Multiflagellate swimming with tilting angle $\theta_0 = 0$, filament number $N = 8$, circle radius $r_0/L = 0.2$, and elastoviscous number $\eta=3$. The actuation amplitude $\theta_{\mathrm{A}} = 0.3$. In Video 1, $\Delta\phi = 0$, and in Video 2, $\Delta\phi = \pi$. The efficiency for the swimmer in Video 1 is $\mathcal{E} = 0.37\%$, and the efficiency in Video 2 is $\mathcal{E} = 0.17\%$.

Videos 3 and 4: Multiflagellate swimming with $\theta_0 = 1.0$ and $\theta_{\mathrm{A}} = 1.0$. Other parameters are the same as Videos 1 and 2. In Video 3, $\Delta\phi = 0$, and in Video 4, $\Delta\phi = \pi$. The efficiency for the swimmer in Video 3 is $\mathcal{E} = 0.14\%$, and the efficiency in Video 4 is $\mathcal{E} = 0.39\%$.

Videos 5 and 6: Multiflagellate swimming a nonzero intrinsic curvature $\kappa_0 = -1.0$. Other parameters are the same as Videos 3 and 4. The efficiency for the swimmer in Video 5 is $\mathcal{E} = 1.1\%$, and the efficiency in Video 6 is $\mathcal{E} = 1.5\%$.

Videos 7 and 8: Multiflagellate swimming with a nonzero intrinsic curvature $\kappa_0 = -1.0$ and $N = 16$. Other parameters are the same as Videos 3 and 4. The efficiency for the swimmer in Video 7 is $\mathcal{E} = 1.4\%$, and the efficiency in Video 8 is $\mathcal{E} = 2.2\%$. The time-averaged swimming speed for the swimmer in Video 8 is $\langle U \rangle = 0.19$ $L/\tau$. Using the dimensional parameters in Table~\ref{table1}, this corresponds to $142.5$ $\mu$m/s or around 10 filament lengths per second, which is comparable to the speed of \textit{P. octopus} measured in experiments~\cite{Hori87}.

\section{Boundary conditions and numerical methods}
\subsection{Filament dynamics}
From Johnson's nonlocal slender-body theory, the filaments are govern by the coupled equations,
\begin{equation}\label{slender_body}
8\pi\mu \left(\frac{\partial \textbf{r}_j}{\partial t} - \textbf{v}_j \right) = \Lambda[\textbf{f}_j] + K[\textbf{f}_j].
\end{equation}
The filament force per unit length $\textbf{f}_j$ is given by the Euler-Bernoulli elasticity,
\begin{equation}
\textbf{f}_j = -B\frac{\partial^4 \textbf{r}_j}{\partial s^4} - B\kappa_0\frac{\partial}{\partial s}(\kappa \textbf{p}_j) + \frac{\partial}{\partial s}(T\textbf{p}_j),
\end{equation}
where $\kappa_0$ is the intrinsic curvature and the local curvature $\kappa = \partial \theta (s,t)/\partial s$ with $\theta(s,t)$ the tangent angle. In this work, we mainly focus on the case of $\kappa_0 = 0$. The operator $\Lambda[\textbf{f}_j]$ in Eq.~(\ref{slender_body}) captures the local drag anisotropy, 
\begin{equation}
\Lambda[\textbf{f}_j] = [c(\textbf{I}+\textbf{p}_j \textbf{p}_j)+2(\textbf{I}-\textbf{p}_j \textbf{p}_j)] \cdot \textbf{f}_j,
\end{equation}
where $c=|\ln(\epsilon^2 e)|$ and the aspect ratio $\epsilon=a/L$. The nonlocal operator $K$ in Eq.~(\ref{slender_body}) is given by
\begin{equation}
K[\textbf{f}_j](s) = \int_0^1 \left[\frac{\textbf{I}+\hat{\textbf{R}}_{jj}\hat{\textbf{R}}_{jj}}{|\textbf{R}_{jj}(s,s')|}\cdot \textbf{f}_j(s') -\frac{\textbf{I}+\textbf{p}_j(s)\textbf{p}_j(s)}{|s-s'|} \cdot\textbf{f}_j(s) \right]\,ds',
\end{equation}
where $\textbf{R}_{jj}(s,s') = \textbf{r}_j(s) - \textbf{r}_j(s')$ and $\hat{\textbf{R}}_{jj} = \textbf{R}_{jj}/|\textbf{R}_{jj}|$. The disturbance velocity $\textbf{v}_j$ at filament $j$ includes the nonlocal Stokeslet flows from all other filaments,
\begin{equation}
\textbf{v}_j(s) = \sum_{k\ne j}^N\textbf{v}_{k\to j} = \frac{1}{8\pi\mu}\sum_{k \ne j}^N \int_{0}^{L}\frac{\textbf{I} + \hat{\textbf{R}}_{jk}\hat{\textbf{R}}_{jk}}{|\textbf{R}_{jk}(s,s')|}\cdot \textbf{f}_k(s') \,ds',
\end{equation}
where $\textbf{R}_{jk}(s,s') = \textbf{r}_j(s) - \textbf{r}_k(s')$. Below in this section we drop the filament index $j$ and work in non-dimensional form. 

Scaling length with the filament length $L$, time with the actuation period $\tau$, and force with $B/L^2$, Eq.~(\ref{slender_body}) written in non-dimensional form is
\begin{equation}\label{slender_body_nonD}
\eta^4 \left(\frac{\partial \textbf{r}}{\partial t} - \textbf{v} \right) = \Lambda[\textbf{f}] + K[\textbf{f}],
\end{equation}
where the elastoviscous number $\eta = L/(B\tau/8\pi\mu)^{1/4}$. Using realistic parameters summarized in Table~\ref{table1}, we estimate $\eta \approx 3$.
\begin{table}[t]
\centering
\begin{tabular}{ c|c|c}
 parameter & symbol & value \\ 
 \hline
 flagellum length & $L$ & 15 $\mu$m~\cite{Wan18} \\
 flagellum aspect ratio & $\epsilon$ & 1/50 \\
 beating frequency & $1/\tau$ & 50 Hz~\cite{Quaranta15} \\
 bending rigidity & $B$ & 840 pN$\cdot$$\mu$m$^2$~\cite{Xu16} \\
 fluid viscosity & $\mu$ & $10^3$ Pa$\cdot$s \\
 body radius & $b$ (equatorial), $l$ (polar) & 8, 15 $\mu$m~\cite{Wan18} 
\end{tabular}
\caption{Dimensional physical parameters of natural flagella. The elastoviscous number is estimated as $\eta = L/(B\tau/8\pi\mu)^{1/4} \approx 3.0$. The velocity scale $L/\tau \approx 750$ $\mu$m/s.}
\label{table1}
\end{table}

The filament's inextensibility condition requires that
\begin{equation}
\frac{\partial}{\partial t}\left(\frac{\partial \textbf{r}}{\partial s} \cdot \frac{\partial \textbf{r}}{\partial s}\right) = 0 \Rightarrow \frac{\partial \textbf{r}}{\partial s} \cdot \frac{\partial^2 \textbf{r}}{\partial s\partial t} = 0.
\end{equation}
Differentiating $\partial\textbf{r}/\partial t$ with respect to the arc length and taking the tangential component, we obtain the tension equation,
\begin{equation}\label{tension}
\begin{aligned}
2c\frac{\partial^2 T}{\partial s^2} - (c+2)\left(\frac{\partial \theta}{\partial s}\right)^2 T &= -\eta^4\frac{\partial\textbf{v}}{\partial s}\cdot \textbf{p} - \frac{\partial K[\textbf{f}]}{\partial s}\cdot \textbf{p} - 6c\left(\frac{\partial^2 \theta}{\partial s^2}\right)^2 - (7c+2)\frac{\partial\theta}{\partial s}\frac{\partial^3\theta}{\partial s^3} \\
&\quad + (c+2)\left(\frac{\partial \theta}{\partial s}\right)^4 - (c+2)\kappa_0 \left(\frac{\partial \theta}{\partial s}\right)^3 + 2c\kappa_0\frac{\partial^3 \theta}{\partial s^3},
\end{aligned}
\end{equation}
The normal component of $\partial^2\textbf{r}/\partial s\partial t $ gives the equation of $\theta$,
\begin{equation}\label{theta}
\begin{aligned}
\eta^4\frac{\partial \theta}{\partial t} + (c+2)\frac{\partial^4\theta}{\partial s^4} &= \eta^4 \frac{\partial \textbf{v}}{\partial s} \cdot \textbf{p}^{\perp} + \frac{\partial K[\textbf{f}]}{\partial s}\cdot \textbf{p}^{\perp} + (9c+6)\left(\frac{\partial \theta}{\partial s}\right)^2\frac{\partial^2\theta}{\partial s^2} + (3c+2)\frac{\partial T}{\partial s}\frac{\partial \theta}{\partial s} \\
&\quad + (c+2)T\frac{\partial^2\theta}{\partial s^2} - (4c+4)\kappa_0 \frac{\partial \theta}{\partial s}\frac{\partial^2\theta}{\partial s^2}.
\end{aligned}
\end{equation}

\subsection{Boundary conditions}\label{boundary_condition}
We first consider the synchronous gait. The rotational symmetry of the model can be used to reduce the computational cost, since the motions of all filaments are identical. We only need to solve the dynamics of one arbitrary filament. At each time instant, the positions, velocities, and forces of other filaments can be obtained by performing rotations of that filament around the central axis. 

Denote the unit vector that lies in the filament's beating plane and perpendicular to $\hat{\textbf{e}}_1$ as $\hat{\textbf{e}}_1^{\perp}$, which is related with $\hat{\textbf{e}}_2$ and $\hat{\textbf{e}}_3$ via $\hat{\textbf{e}}_1^{\perp} = \hat{\textbf{e}}_2 \cos\beta + \hat{\textbf{e}}_3\sin\beta$. From Eq.~(\ref{slender_body_nonD}), the filament velocity $\partial\textbf{r}/\partial t = \textbf{v} + \eta^{-4}K[\textbf{f}] + \eta^{-4}\Lambda [\textbf{f}]$. At $s=0$, we require that the filament velocity along $\hat{\textbf{e}}_1^{\perp}$ is zero, $\hat{\textbf{e}}_1^{\perp} \cdot \partial \textbf{r}/\partial t = 0$. Rewriting the local operator in terms of $\theta$ and $T$, this condition leads to
\begin{equation}\label{yt_zero}
\begin{aligned}
\eta^{-4}\left(6c\frac{\partial\theta}{\partial s}\frac{\partial^2\theta}{\partial s^2} + 2c\frac{\partial T}{\partial s} - 2c\kappa_0\frac{\partial^2\theta}{\partial s^2}\right)\sin\theta + \eta^{-4}(c+2)\left\{-\left[\frac{\partial^3 \theta}{\partial s^3}-\left(\frac{\partial\theta}{\partial s}\right)^3\right] + \frac{\partial\theta}{\partial s} T -\kappa_0\left(\frac{\partial\theta}{\partial s}\right)^2\right\} \cos\theta \\
+ \left[(\textbf{v}+ \eta^{-4} K)\cdot \hat{\textbf{e}}_2 \cos\beta + (\textbf{v} + \eta^{-4} K)\cdot \hat{\textbf{e}}_3 \sin\beta \right] = 0.
\end{aligned}
\end{equation}
The force-free condition at $s=0$ along the $e_1$ direction is given by
\begin{equation}\label{force_free1}
\left(-\frac{\partial^3\textbf{r}}{\partial s^3}-\kappa_0\kappa\textbf{p}+T\textbf{p}\right) \cdot \hat{\textbf{e}}_1 = 0 \Rightarrow \frac{\partial^2 \theta}{\partial s^2}\sin\theta + \left(\frac{\partial\theta}{\partial s}\right)^2\cos\theta - \kappa_0\frac{\partial\theta}{\partial s}\cos\theta + T\cos\theta = 0.
\end{equation}
To close the system, the force-free and torque-free conditions at $s=1$ are
\begin{equation}\label{free_end}
\theta_s = \kappa_0,\quad \theta_{ss} = 0,\quad T = 0.
\end{equation}

In the asynchronous gait, neighboring filaments are beating with a phase difference. We need to solve two sets of equations for two arbitrary filaments with different phases. Equations~(\ref{tension}), (\ref{theta}), (\ref{yt_zero}), and (\ref{free_end}) still hold for each filament. The force-free condition Eq.~(\ref{force_free1}) needs to be expressed using the sum of filament forces,
\begin{equation}\label{force_free2}
\sum_{j=1}^2 \left[\frac{\partial^2 \theta_j}{\partial s^2}\sin\theta_j + \left(\frac{\partial\theta_j}{\partial s}\right)^2\cos\theta_j - \kappa_0\frac{\partial\theta_j}{\partial s}\cos\theta_j + T_j\cos\theta_j\right] = 0.
\end{equation}
where the subscript integer $j$ denotes the two filaments. 

The final boundary condition for the asynchronous gait at $s=0$ is that the $e_1$-component velocities of the two filaments are the same,
\begin{equation}\label{equal_e1_velocity}
\begin{aligned}
\eta^{-4}\left(6c\frac{\partial \theta_1}{\partial s}\frac{\partial^2\theta_1}{\partial s^2} + 2c\frac{\partial T_1}{\partial s} - 2c\kappa_0\frac{\partial^2 \theta_1}{\partial s^2}\right)\cos\theta_1 \\
- \eta^{-4}(c+2)\left\{-\left[\frac{\partial^3\theta_1}{\partial s^3}-\left(\frac{\partial\theta_1}{\partial s}\right)^3\right] + \frac{\partial\theta_1}{\partial s} T_1 -\kappa_0\left(\frac{\partial\theta_1}{\partial s}\right)^2\right\}\sin\theta_1 + (\textbf{v}_1 +\eta^{-4}K)\cdot\hat{\textbf{e}}_1 = \\
\eta^{-4}\left(6c\frac{\partial \theta_2}{\partial s}\frac{\partial^2\theta_2}{\partial s^2} + 2c\frac{\partial T_2}{\partial s} - 2c\kappa_0\frac{\partial^2 \theta_2}{\partial s^2}\right)\cos\theta_2 \\
- \eta^{-4}(c+2)\left\{-\left[\frac{\partial^3\theta_2}{\partial s^3} -\left(\frac{\partial\theta_2}{\partial s}\right)^3\right] + \frac{\partial\theta_2}{\partial s} T_2 -\kappa_0\left(\frac{\partial\theta_2}{\partial s}\right)^2\right\}\sin\theta_2 
 + (\textbf{v}_2 + \eta^{-4}K)\cdot\hat{\textbf{e}}_1.
\end{aligned}
\end{equation}

\begin{figure}[t]
\centering
\includegraphics[bb= 0 5 365 145, scale=0.8,draft=false]{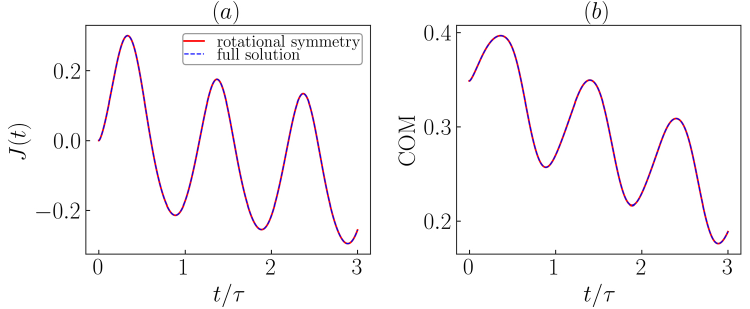}
\caption{The results of the numerical solution strategy utilizing the rotational symmetry are the same as the results of the full solution by evolving all filaments. (a) The position of the hinged point $J(t)$ and (b) the position of the center-of-mass (COM) location as a function of time.}
\label{fig_S1}
\end{figure}

We have also tried to evolve the dynamics of all filaments simultaneously, but with a much greater computational cost. The results of the two approaches are the same, as shown in Fig.~\ref{fig_S1}. 

\subsection{Numerical methods}
The governing equations are solved using a second-order finite-difference scheme. Our goal is to compute $\theta^{n+1}$ and $T^{n+1}$ given $\theta^n$ and $T^n$. Discretize the arc length as $s_m = m\Delta s$ with $m = 0$, 1, $\cdots$, $1/\Delta s$. Discretize time as $t_n = n\Delta t$ and denote with a superscript $n$ the quantities at time step $t_n$. Schematically, we write the $\theta$ equation as
\begin{equation}\label{theta_schematic}
\begin{aligned}
\frac{3\theta^{n+1}-4\theta^n+\theta^{n-1}}{2\Delta t} + \eta^{-4}(c+2) \frac{\partial^4\theta^{n+1}}{\partial s^4} = \left(\textbf{v}_s \cdot \textbf{p}^{\perp} + \eta^{-4}\frac{\partial K}{\partial s}\cdot \textbf{p}^{\perp}\right)^{n} + g^{n, n-1}.
\end{aligned}
\end{equation}
where a second-order backward scheme is used to approximate the time derivative. The fourth-order term is evaluated implicitly at time step $n+1$. Other terms, included in $g^{n,n-1}$, are extrapolated using values from time steps $n$ and $n-1$. 

Due to the nonlinear boundary conditions, we use Newton's method~\cite{Hu23}. Denote the solutions at the $k$-th Newton's iteration with a superscript $k$. We linearize (\ref{theta_schematic}) with $\theta^{n+1} = \theta^k + \delta\theta$,  
\begin{equation}\label{linear_theta}
\begin{aligned}
\delta\theta + \frac{2\Delta t \eta^{-4} (c+2)}{3}\frac{\partial^4 \delta\theta}{\partial s^4} = \frac{4}{3}\theta^n - \frac{1}{3}\theta^{n-1} + \frac{2\Delta t}{3}\left(\textbf{v}_s\cdot\textbf{p}^{\perp} + \eta^{-4}\frac{\partial K}{\partial s}\cdot \textbf{p}^{\perp}\right)^n + g^{n,n-1} - \eta^{-4}(c+2)\frac{\partial^4\theta^{k}}{\partial s^4}.
\end{aligned}
\end{equation}
The tension equation is linearized with $T^{n+1} = T^k + \delta T$,
\begin{equation}\label{linear_tension}
\begin{aligned}
2c\frac{\partial^2 \delta T}{\partial s^2} - (c+2)\left(\frac{\partial \theta^k}{\partial s} \right)^2\delta T + \left[(7c+2)\frac{\partial \theta^k}{\partial s} - 2c\kappa_0\right]\frac{\partial^3\delta\theta}{\partial s^3} + 12c\frac{\partial^2 \theta^k}{\partial s^2}\frac{\partial^2\delta\theta}{\partial s^2} \\
+ \left[-2(c+2)\frac{\partial \theta^k}{\partial s} T^k +(7c+2)\frac{\partial^3\theta^k}{\partial s^3} - 4(c+2)\left(\frac{\partial\theta^k}{\partial s}\right)^3 + 3(c+2)\kappa_0\left(\frac{\partial\theta^k}{\partial s}\right)^2\right]\frac{\partial \delta\theta}{\partial s} \\
= \left(-\eta^4\frac{\partial\textbf{v}}{\partial s}\cdot \textbf{p} - \frac{\partial K[\textbf{f}]}{\partial s}\cdot \textbf{p}\right)^n + M[\theta^k, T^k],
\end{aligned}
\end{equation}
where $M[\theta^k, T^k]$ collects terms evaluated at iteration $k$. The boundary conditions are linearized in a similar way. Results from previous time step are used as the initial guesses. Solving the resulting linear system for $\delta\theta$ and $\delta T$ and iterating until converge, we obtain $\theta^{n+1}$ and $T^{n+1}$. In our simulations, we set $c = 15$, $\Delta s = 10^{-2}$, and $\Delta t = 10^{-4}$--$10^{-3}$.

\section{Hydrodynamic coupling strength for parallel filaments}

We consider initially parallel filaments with $\theta_0 = 0$. At the limit of small actuation amplitude $\theta_{\mathrm{A}} \ll 1$, the arc length $s\approx x$. We assume that the relative distances between filaments are not changing with time, but the filament forces can be time-dependent. The goal is to approximate the nonlocal integral over the Stokeslet as a local velocity-force relation, in which the constant prefactor represents the strength of the hydrodynamic interactions, i.e., the hydrodynamic coupling strength. 

The velocity generated by filament $k$ at filament $j$ is computed by integrating the distributed Stokeslet along the filament arc length,
\begin{equation}
\textbf{v}_{k \to j}(x) = \frac{1}{8\pi\mu}\int_{0}^{L}\frac{\textbf{I}+\hat{\textbf{R}}_{jk}\hat{\textbf{R}}_{jk}}{R_{jk}(x, x')}\cdot \textbf{f}_k(x')\,dx',
\end{equation}
where $\textbf{R}_{jk}(x,x') = \textbf{r}_j(x)-\textbf{r}_k(x') = [x-x', r_0(\cos\beta_j-\cos\beta_k), r_0(\sin\beta_j-\sin\beta_k)]$. The magnitude $R_{jk}(x, x') = [(x-x')^2+2(1-\lambda_{jk}) r_0^2]^{1/2}$, where we define $\lambda_{jk} = \cos(\beta_j-\beta_k)$. We scale length with $L$, force with $B/L^2$, and velocity with $L/\tau$. The non-dimensional velocity is 
\begin{equation}
\textbf{v}_{k \to j}(x) = \frac{1}{\eta^4}\int_{0}^{1}\frac{\textbf{I}+\hat{\textbf{R}}_{jk}\hat{\textbf{R}}_{jk}}{R_{jk}(x, x')}\cdot \textbf{f}_k(x')\,dx'.
\end{equation}
The filament force per unit length is (in $\{e_1,e_2,e_3\}$ frame) 
\begin{equation}
\textbf{f}_k(x) = \left(0, f_k\cos\beta_k, f_k\sin\beta_k \right),
\end{equation}
where the bending force $f_k(x) = -\partial^4 y_k/\partial x^4$. The $e_2$- and $e_3$-components of the induced velocity are
\begin{equation}
\begin{aligned}
v_{e_2, k \to j} &= -\frac{r_0^2(\cos\beta_j-\cos\beta_k)(1-\lambda_{jk})}{\eta^4}\int_0^1 \frac{1}{R_{jk}^3}f_k(x') \,dx' + \frac{\cos\beta_k}{\eta^4}\int_0^1 \frac{1}{R_{jk}} f_k(x')\,dx', \\
v_{e_3, k \to j} &= -\frac{r_0^2(\sin\beta_j-\sin\beta_k)(1-\lambda_{jk})}{\eta^4}\int_0^1 \frac{1}{R_{jk}^3} f_k(x') \,dx' + \frac{\sin\beta_k}{\eta^4}\int_0^1 \frac{1}{R_{jk}} f_k(x')\,dx'.
\end{aligned}
\end{equation}
Since the filament undergoes planar motion, we take the projection of the induced velocities onto the filament's beating plane, 
\begin{equation}
v^{\perp}_{k \to j}(x) = v_{e_2, k \to j}\cos\beta_j + v_{e_3, k \to j}\sin\beta_j = \frac{1}{\eta^4}\int_0^1 \frac{\lambda_{jk} R_{jk}^2 - r_0^2(1-\lambda_{jk})^2}{R_{jk}^3} f_k (x')\, dx'.
\end{equation}
This integral can be separated into three parts by introducing an intermediate scale $\delta$ with $r_0\sqrt{2(1-\lambda_{jk})} \ll \delta \ll 1$, 
\begin{equation}
\frac{1}{\eta^4}\int_0^1 \cdots f_k(x')\,dx' = \frac{1}{\eta^4}\left(\int_0^{x-\delta} + \int_{x+\delta}^1 + \int_{x-\delta}^{x+\delta}\right) \cdots f_k(x') \,dx',
\end{equation}
where the first two terms are nonlocal parts and the last term is the local part. For the nonlocal parts, $R_{jk} \approx \sqrt{(x-x')^2} \gg r_0\sqrt{(1-\lambda_{jk})}$. Therefore, the nonlocal parts can be approximated as,
\begin{equation}
\frac{\lambda_{jk}}{\eta^4} \left[\int_0^{x-\delta} \frac{1}{x-x'} f_k(x') \,dx' + \int_{x+\delta}^{1} \frac{1}{x'-x} f_k(x') \,dx' \right] \approx -2\lambda_{jk} \ln\delta \frac{1}{\eta^4} f_k(x).
\end{equation}
To calculate the local part, we perform the change of variable $\Delta = x'-x$. The local part can then be calculated as
\begin{equation}
\begin{aligned}
\frac{1}{\eta^4}\int_{-\delta}^{\delta} \frac{\lambda_{jk}\Delta^2+r_0^2\left[2\lambda_{jk}(1-\lambda_{jk})-(1-\lambda_{jk})^2\right]}{\left[\Delta^2+2(1-\lambda_{jk}) r_0^2\right]^{\frac{3}{2}}} f_k\,d\Delta \\
\approx \left\{-1+\lambda_{jk}+\lambda_{jk}\ln 4-\lambda_{jk}\ln\left[2(1-\lambda_{jk})r_0^2\right] + 2\lambda_{jk}\ln \delta\right\} \frac{1}{\eta^4}f_k(x).
\end{aligned}
\end{equation}
Summing up the local part and nonlocal parts, the $\ln\delta$ terms cancel out. The final expression for the induced velocity is independent of the intermediate length $\delta$,
\begin{equation}
v^{\perp}_{k \to j}(x) = 2\ln(1/r_0) \Gamma_{jk} \frac{1}{\eta^4}f_k(x),
\end{equation}
where we define the coupling strength between filament $j$ and $k$ as
\begin{equation}
\Gamma_{jk} = \frac{1}{2\ln r_0} \left[1+\lambda_{jk}\ln\left(\frac{1-\lambda_{jk}}{2 e}r_0^2\right)\right].
\end{equation}
Restoring the dimensions,
\begin{gather}
v_{k\to j}(x) = \ln(L/r_0)\Gamma_{jk}\frac{1}{4\pi\mu}f_k(x), \\
\Gamma_{jk} = \frac{1}{2\ln (r_0/L)} \left[1+\lambda_{jk}\ln\left(\frac{1-\lambda_{jk}}{2 e}\frac{r_0^2}{L^2}\right)\right]. \label{gamma_jk}
\end{gather}
The above two equations are given in the main text.

\section{Small-amplitude solution for parallel filaments}

To develop the small-amplitude solution, we first neglect the nonlocal term $K$, since it is smaller than the local term $\Lambda$ approximately by a factor of $1/\ln(L/a)$. We focus on intrinsically straight filaments with $\kappa_0 = 0$. Equation~(\ref{slender_body}) can be rewritten as
\begin{equation}\label{slender_body2}
\left[\xi_{\parallel}\textbf{p}_j\textbf{p}_j + \xi_{\perp}(\textbf{I}-\textbf{p}_j\textbf{p}_j)\right]\cdot \left(\frac{\partial \textbf{r}_j}{\partial t} - \textbf{v}_j\right) = -B\frac{\partial^4 \textbf{r}_j}{\partial s^4} + \frac{\partial}{\partial s}\left(T_j\frac{\partial \textbf{r}_j}{\partial s}\right),
\end{equation}
where the parallel and perpendicular drag coefficients are
\begin{equation}
\begin{aligned}
\xi_{\parallel} = \frac{4\pi\mu}{c} = \frac{4\pi\mu}{2\ln(L/a) - 1}, \quad \xi_{\perp} = \frac{8\pi\mu}{c+2} = \frac{8\pi\mu}{2\ln(L/a) +1}.
\end{aligned}
\end{equation}

For small driving amplitude $\theta_{\mathrm{A}}$, the filament deformation is small $y\sim \theta_{\mathrm{A}} L$, and the arc length $s \approx x$. Since the filament motion consists of a translation and a planar beating motion, the filament velocity can be written as $\partial\textbf{r}_j(s,t)/\partial t \approx U(t)\hat{\textbf{e}}_1 + \partial y_j(x,t)/\partial t \hat{\textbf{y}}_j$. The tangent vector $\textbf{p}_j \approx (1, \partial y_j/\partial x)$. 

From Eq.~(\ref{tension}), we can see that $T \sim \theta_{\mathrm{A}}^2$. Keeping terms up to $\mathcal{O}(\theta_{\mathrm{A}}^2)$, Eq.~(\ref{slender_body2}) written in $\{x, y\}$ frame becomes~\cite{Lauga07},
\begin{gather}
\xi_{\parallel} U(t) + \left(\xi_{\parallel}-\xi_{\perp}\right)\frac{\partial y_j}{\partial x}\left(\frac{\partial y_j}{\partial t}-v_j\right) = \frac{\partial T_j}{\partial x}, \label{x_component} \\
\xi_{\perp}\left(\frac{\partial y_j}{\partial t}-v_j\right) = -B\frac{\partial^4 y_j}{\partial x^4}, \label{y_component}
\end{gather}
Equation~(\ref{y_component}) is presented in the main text. The boundary conditions of $y_j$ are
\begin{gather}
y_j(0, t) = 0,\quad \frac{\partial y_j}{\partial x}(0, t) = \theta_{\mathrm{A}} \sin(2\pi t/\tau + \phi_j),\\
\frac{\partial^2 y_j}{\partial x^2}(L, t) = 0,\quad \frac{\partial^3 y_j}{\partial x^3}(L, t) = 0.
\end{gather}
The force-free condition requires the tension of each filament to be zero at $x=L$ and the total tension to be zero at $x = 0$, $\sum_{j=1}^{N} T_j (x = 0, t) = 0$. After integration by parts of Eq.~(\ref{x_component}) along the arc length and eliminating $T_j$ with the boundary conditions, we obtain an expression for $U$ in terms of the boundary values of the derivatives of $y_j$,
\begin{equation}\label{coupled_U}
\begin{aligned}
U(t) = -\frac{(\gamma-1)B}{N L \xi_{\perp}} \sum_{j=1}^N \left[\frac{1}{2}\left(\frac{\partial^2 y_j}{\partial x^2}\right)^2-\frac{\partial y_j}{\partial x}\frac{\partial^3 y_j}{\partial x^3}\right]_{x=0}.
\end{aligned}
\end{equation}
We define the drag coefficient ratio $\gamma = \xi_{\perp}/\xi_{\parallel} = 2c/(c+2)$. For extremely slender filament $a/L \ll 1$, $\gamma \approx 2$. 

Due to the symmetry of the system, the filament motions are essentially identical except that they may have different phases. Therefore, the time-averaged speed of the swimmer can be written in terms of an arbitrary filament, 
\begin{equation}\label{time_averaged_U}
\begin{aligned}
\langle U\rangle = \frac{1}{\tau}\left|\int_0^\tau U(t) \,dt\right| = \frac{(\gamma-1)B}{L \xi_{\perp}} \left\langle \left[\frac{1}{2}\left(\frac{\partial^2 y_j}{\partial x^2}\right)^2-\frac{\partial y_j}{\partial x}\frac{\partial^3 y_j}{\partial x^3}\right]_{x=0}\right\rangle.
\end{aligned}
\end{equation}

If the filaments are loaded with a sphere of radius $b$. The total filament force needs to balance the viscous drag of the sphere. Denote the speed of the load swimmer as $U_{\mathrm{L}}$. The tensions satisfy the boundary condition 
\begin{equation}
\sum_{j = 1}^N T_j(x = 0, t) = 6\pi\mu b U_{\mathrm{L}}(t).
\end{equation}
Repeating the above calculation, we obtain
\begin{equation}\label{drag_balance_body}
(N\xi_{\parallel} L + 6\pi\mu b) \langle U_{\mathrm{L}}\rangle = N B\frac{\gamma - 1}{\gamma}\left\langle \left[\frac{1}{2}\left(\frac{\partial^2 y_j}{\partial x^2}\right)^2 - \frac{\partial y_j}{\partial x}\frac{\partial^3 y_j}{\partial x^3} \right]_0 \right\rangle.
\end{equation}
Here, we recognize that the right hand side is the total propulsion generated without the load, i.e., $N \xi_{\parallel} L \langle U\rangle$. The left hand side is the total viscous drag of the filaments and the load when translating together with velocity $\langle U_{\mathrm{L}} \rangle$. Therefore, Eq.~(\ref{drag_balance_body}) can be rewritten as
\begin{equation}\label{drag_balance_body_simple}
(ND_{\mathrm{f}} + 6\pi\mu b) \langle U_{\mathrm{L}}\rangle = N D_{\mathrm{f}} \langle U \rangle,
\end{equation}
where we denote the resistivity of a single filament as $D_{\mathrm{f}} = \xi_{\parallel} L$. Equation~(\ref{drag_balance_body_simple}) is used in the main text. This simple relation is also valid for $\theta_0 > 0$.

\subsection{Solution without HIs}

\begin{figure}[t]
\centering
\includegraphics[bb= 0 5 365 210, scale=0.6,draft=false]{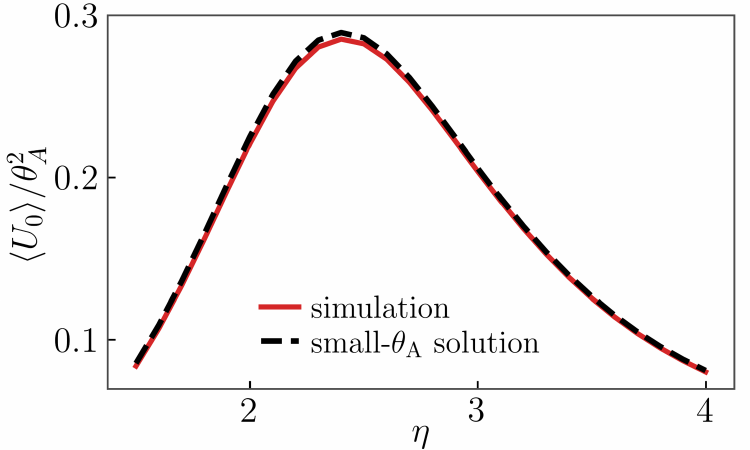}
\caption{The time-averaged velocity without iHIs $\langle U_0 \rangle$ obtained from simulations agrees well with the small-$\theta_{\mathrm{A}}$ solution. Other parameters are $\theta_0 = 0$ and $\theta_{\mathrm{A}} = 0.2$.}
\label{fig_S2}
\end{figure}

We solve Eq.~(\ref{y_component}) in Fourier space by writing $y_0(x,t) = \mathrm{Im}\{\zeta(x) \exp(2\pi i t/\tau)\}$ (the subscript index $j$ is dropped and the subscript 0 denotes solution without iHIs). The complex function $\zeta(x)$ is a solution to the equation
\begin{equation}\label{zeta}
\left[2\pi i \xi_{\perp}/(B\tau)  + \frac{d^4}{d x^4}\right]\zeta(x) = 0.
\end{equation}
At the boundaries, 
\begin{gather}
\zeta(0) = 0,\quad \frac{d \zeta}{d x}(0) = \theta_{\mathrm{A}}, \\
\frac{d^2 \zeta}{d x^2}(L) = 0,\quad \frac{d^3 \zeta}{d x^3}(L) = 0.
\end{gather}
The analytic solution to Eq.~(\ref{zeta}) is given by
\begin{equation}\label{zeta_noHIs}
\zeta (x) = \sum_{m = 0}^{3} A_{m}e^{\alpha_{0,m} x},
\end{equation}
where $\alpha_{0,m} = [2\pi\xi_{\perp}/(B\tau)]^{1/4} e^{(-1/4+m)i\pi/2}$. They satisfy the following relations: $\alpha_{0,1} = i \alpha_{0,0}$, $\alpha_{0,2} = i\alpha_{0,1}$, and $\alpha_{0,3} = i\alpha_{0,2}$. From the boundary conditions, the coefficients $A_{m}$ satisfy the linear system
\begin{equation}\label{A_noHIs}
\begin{bmatrix}
1 & 1 & 1 & 1 \\
\alpha_{0,0} & \alpha_{0,1} & \alpha_{0,2} & \alpha_{0,3}\\
(\alpha_{0,0})^2 e^{\alpha_0 L} & (\alpha_{0,1})^2 e^{\alpha_1 L} & (\alpha_{0,2})^2 e^{\alpha_2 L} & (\alpha_{0,3})^2 e^{\alpha_3 L} \\
(\alpha_{0,0})^3 e^{\alpha_0 L} & (\alpha_{0,1})^3 e^{\alpha_1 L} & (\alpha_{0,2})^3 e^{\alpha_2 L} & (\alpha_{0,3})^3 e^{\alpha_3 L} 
\end{bmatrix}
\begin{bmatrix}
A_0 \\
A_1 \\
A_2 \\
A_3 
\end{bmatrix}
= 
\begin{bmatrix}
0 \\
\theta_{\mathrm{A}} \\
0 \\
0 
\end{bmatrix}.
\end{equation}
From Eq.~(\ref{time_averaged_U}), the time-averaged swimming velocity can be written in terms of $\zeta$,
\begin{equation}\label{U_analytic_zeta}
\begin{aligned}
\langle U\rangle = \frac{(\gamma-1)B}{L \xi_{\perp}} \left[\frac{1}{4}\Big|\frac{d^2 \zeta}{d x^2}(0)\Big|^2 - \frac{1}{2}\theta_{\mathrm{A}}\text{Re}\left\{\frac{d^3\zeta}{d x^3}(0)\right\}\right].
\end{aligned}
\end{equation}
The analytic solution obtained from Eqs.~(\ref{zeta_noHIs})--(\ref{U_analytic_zeta}) is shown in Fig.~\ref{fig_S2}, which is in good agreement with the simulation result. 

For large $\eta$, viscous force dominates, and the filament deformation decays rapidly along the arc length. The boundary conditions at $x=L$ are satisfied automatically. Therefore, only $A_m$ for which $\alpha_{0,m}$ have a negative real part are allowed, i.e., $A_2$ and $A_3$~\cite{Wiggins98_2}. Applying the boundary conditions at $x = 0$, 
\begin{gather}
A_2 + A_3 = 0,\\
\alpha_{0,2} A_2 + \alpha_{0,3} A_3 = \theta_{\mathrm{A}}.
\end{gather}
The solutions are $A_2 = \theta_{\mathrm{A}}/(\alpha_{0,2}-\alpha_{0,3})$ and $A_3 = -A_2$. The complex function $\zeta(x)$ is
\begin{equation}
\zeta(x) \approx A_2 e^{\alpha_{0,2} x} + A_3 e^{\alpha_{0,3} x} = \frac{\theta_{\mathrm{A}}}{(1-i)\alpha_{0,2}}(e^{\alpha_{0,2} x}-e^{i\alpha_{0,2} x}).
\end{equation}
Substituting into Eq.~(\ref{U_analytic_zeta}) and taking $\gamma \approx 2$, we obtain the swimming velocity without HIs,
\begin{equation}
\begin{aligned}
\langle U_0 \rangle &= \theta_{\mathrm{A}}^2\frac{L}{\tau}\frac{(2\pi)^{1/2}(2-\sqrt{2})}{4}\left(\frac{B \tau}{\xi_{\perp} L^4}\right)^{1/2} \\
&\sim \theta_{\mathrm{A}}^2\frac{L}{\tau}\eta^{-2}[\ln(L/a)]^{1/2}.
\end{aligned}
\end{equation}
The above equation is the same as Eq. (67) of~\cite{Lauga07}, taking $\theta_0 = 0$ and $\mathcal{R}_{\parallel}^{FU} = 0$ there.

\subsection{Synchronous beating with HIs}

For the synchronous state with HIs, the deformation satisfies 
\begin{equation}
\xi_{\perp} \frac{\partial y_j}{\partial t} + (1+\Gamma_0\Gamma)B \frac{\partial^4 y_j}{\partial x^4} = 0. 
\end{equation}
The solution has a similar form to $y_{0}(x,t)$, 
\begin{equation}
y_j(x,t) = \mathrm{Im}\{\zeta(x) \exp(2\pi i t/\tau)\} = \mathrm{Im}\left\{\exp(2\pi i t/\tau)\sum_{m=0}^3 A_m e^{\alpha_m x}\right\}.
\end{equation}
Here, the exponent $\alpha_m = [2\pi\xi_{\perp}(1+\Gamma_0\Gamma)^{-1}/(B\tau)]^{1/4} e^{(-1/4+m)i\pi/2}$. 

Taking the limit at large $\eta$, the solution of $\zeta$ is 
\begin{equation}\label{zeta_large_eta}
\zeta(x) = \frac{\theta_{\mathrm{A}}}{(1-i)\alpha_2}(e^{\alpha_2 x}-e^{i\alpha_2 x}),
\end{equation}
Substituting the solution of $\zeta$ into Eq.~(\ref{U_analytic_zeta}), we obtain the time-averaged swimming velocity 
\begin{equation}
\begin{aligned}
\langle U \rangle &= \theta_{\mathrm{A}}^2\frac{L}{\tau}\frac{(2\pi)^{1/2}(2-\sqrt{2})}{4}\left[\frac{B \tau}{\xi_{\perp} L^4 (1+\Gamma_0\Gamma)}\right]^{1/2} \\
&= \langle U_0\rangle (1+\Gamma_0\Gamma)^{-1/2}.
\end{aligned}
\end{equation}
The time-averaged rate of work done by the actuator against the viscous fluid is 
\begin{equation}\label{dissipation}
\begin{aligned}
P &= -\frac{B}{\tau} \int_0^{\tau} \int_0^L \frac{\partial y_j}{\partial t} \frac{\partial^4 y_j}{\partial x^4}\, dt\,dx \\
&= (1+\Gamma_0\Gamma)\frac{B^2}{\xi_{\perp}\tau} \int_0^L\int_0^{\tau}\left(\frac{\partial^4 y_j}{\partial x^4}\right)^2\,dt\,dx \\
&= (1+\Gamma_0\Gamma)\frac{B^2}{2\xi_{\perp}} \int_0^L\Big|\frac{d^4 \zeta}{d x^4}\Big|^2\,dx. 
\end{aligned}
\end{equation}
Substituting Eq.~(\ref{zeta_large_eta}) into Eq.~(\ref{dissipation}), we obtain 
\begin{equation}
P \sim \frac{\theta_\mathrm{A}^2 \eta B}{\tau L}(1+\Gamma_0\Gamma)^{-1/4}[\ln (L/a)]^{-1/4}.
\end{equation}

\subsection{Asynchronous beating with HIs}

For the asynchronous gait, the dynamics of the filaments are coupled via interflagella hydrodynamics interactions. Different from the synchronous case, the dynamics of two neighboring filaments with different phases need to be solved simultaneously (taking 1 and 2 for example),
\begin{equation}\label{coupled_yt_phaselag}
\begin{aligned}
\xi_{\perp}\frac{\partial y_1}{\partial t} + (1+\Gamma_0\Gamma_{\mathrm{s}})B\frac{\partial^4 y_1}{\partial x^4} + \Gamma_0\Gamma_{\mathrm{as}}B\frac{\partial^4 y_2}{\partial x^4} = 0, \\
\xi_{\perp}\frac{\partial y_2}{\partial t} + (1+\Gamma_0\Gamma_{\mathrm{s}})B\frac{\partial^4 y_2}{\partial x^4} + \Gamma_0\Gamma_{\mathrm{as}}B\frac{\partial^4 y_1}{\partial x^4} = 0,
\end{aligned}
\end{equation}
where $\Gamma_{\mathrm{s}}$ is the CS by the filaments in the synchronous group with filament 1 or 2 and $\Gamma_{\mathrm{as}}$ is the CS by the filaments in the asynchronous group. For filament 1,
\begin{equation}
\Gamma_{\mathrm{s}} = \sum_{\textrm{odd }j, j \neq 1}^{N} \Gamma_{j1},\quad \Gamma_{\mathrm{as}} = \sum_{\textrm{even }j}^{N} \Gamma_{j1}.
\end{equation}
For filament 2,
\begin{equation}
\Gamma_{\mathrm{s}} = \sum_{\textrm{even }j, j \neq 2}^{N} \Gamma_{j2},\quad \Gamma_{\mathrm{as}} = \sum_{\textrm{odd }j}^{N} \Gamma_{j2}.
\end{equation}
The values of $\Gamma_{\textrm{s}}$ and $\Gamma_{\textrm{as}}$ are the same for filaments 1 and 2. They also satisfy $\Gamma_{\textrm{s}} + \Gamma_{\textrm{as}} = \Gamma$.

The boundary actuations are 
\begin{equation}
\frac{\partial y_1}{\partial x}(0, t) = \theta_{\mathrm{A}} \sin(2\pi t/\tau), \quad \frac{\partial y_2}{\partial x} (0, t) = \theta_{\mathrm{A}} \sin(2\pi t/\tau + \Delta\phi). 
\end{equation}

\begin{figure}[t]
\centering
\includegraphics[bb= 0 5 365 210, scale=0.6,draft=false]{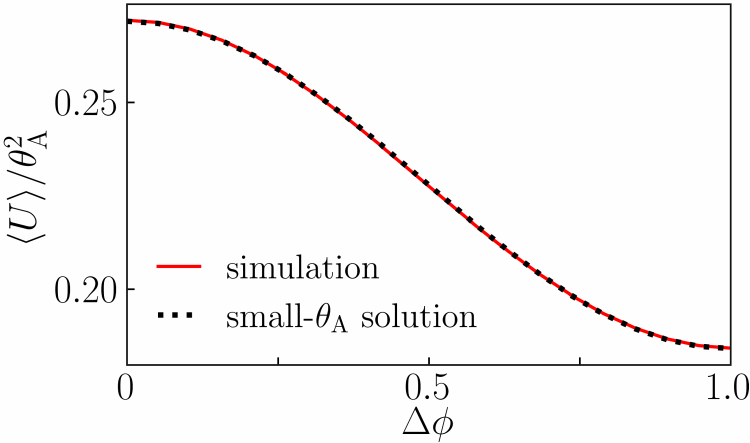}
\caption{The time-averaged swimming speed $\langle U\rangle$ as a function of $\Delta\phi$ for $\theta_0 = 0$, $\theta_{\mathrm{A}} = 0.1$, and $r_0 = 0.2$. The dotted line is the fitting of Eq.~(\ref{average_U_para_phaselag}). The best-fit values of the coupling strength are $\Gamma_{\mathrm{s}}=-1.9$ and $\Gamma_{\mathrm{as}}=-1.1$. The coupling strength estimated using Eq. (3) of the main text [Eq.~(\ref{gamma_jk})] are $\Gamma_{\mathrm{s}}=-2.2$ and $\Gamma_{\mathrm{as}}=-0.5$. The computed values are close to the beat-fit values.}
\label{fig_S3}
\end{figure}
Equation~\ref{coupled_yt_phaselag} can be solved in Fourier space by writing $y_j(x,t) = \text{Im}\{\zeta_j(x) \exp(2\pi i t/\tau)\}$. The complex functions $\zeta_1$ and $\zeta_2$ satisfy 
\begin{equation}\label{coupled_zeta}
\begin{aligned}
2\pi i \frac{\xi_{\perp}}{B\tau} \zeta_1 + (1+\Gamma_0\Gamma_{\mathrm{s}})\frac{d^4 \zeta_1}{dx^4} + \Gamma_0\Gamma_{\mathrm{as}}\frac{d^4 \zeta_2}{dx^4} = 0, \\
2\pi i \frac{\xi_{\perp}}{B\tau} \zeta_2 + (1+\Gamma_0\Gamma_{\mathrm{s}})\frac{d^4 \zeta_2}{dx^4} + \Gamma_0\Gamma_{\mathrm{as}}\frac{d^4 \zeta_1}{dx^4} = 0.
\end{aligned}
\end{equation}
The boundary conditions are
\begin{equation}
\begin{aligned}
\zeta_1(0) = 0,\quad \frac{d \zeta_1}{dx}(0) = \theta_{\mathrm{A}},\quad \frac{d^2 \zeta_1}{d x^2}(1) = 0,\quad \frac{d^3\zeta_1}{d x^3}(1) = 0, \\
\zeta_2(0) = 0,\quad \frac{d \zeta_2}{dx}(0) = \theta_{\mathrm{A}} e^{i\Delta\phi},\quad \frac{d^2 \zeta_2}{d x^2}(1) = 0,\quad \frac{d^3\zeta_2}{d x^3}(1) = 0.
\end{aligned}
\end{equation}
The above two equations can be decoupled by rewriting in terms of $\zeta_1 + \zeta_2$ and $\zeta_1 - \zeta_2$. We define 
\begin{equation}\label{variable_change}
\zeta_{\mathrm{p}} = \zeta_1 + \zeta_2, \quad \zeta_{\mathrm{m}} = \zeta_1 - \zeta_2.
\end{equation}
They satisfy
\begin{gather}
2\pi i \frac{\xi_{\perp}}{B\tau} \zeta_{\mathrm{p}} + (1+\Gamma_0\Gamma)\frac{d^4\zeta_{\mathrm{p}}}{d x^4} = 0, \\
2\pi i \frac{\xi_{\perp}}{B\tau} \zeta_{\mathrm{m}} + (1+\Gamma_0\Gamma_{\mathrm{s}}-\Gamma_0\Gamma_{\mathrm{as}})\frac{d^4\zeta_{\mathrm{m}}}{d x^4} = 0.
\end{gather}
The solutions for $\zeta_{\mathrm{p}}$ and $\zeta_{\mathrm{m}}$ have similar forms to Eq.~(\ref{zeta_noHIs}). At large $\eta$, 
\begin{equation}
\begin{aligned}
\zeta_{\mathrm{p}} = \frac{\theta_{\mathrm{A}}(1+e^{i\Delta\phi})}{(1-i)\alpha_{2}}\left(e^{\alpha_{2}x} - e^{i\alpha_{2}x}\right), \\
\zeta_{\mathrm{m}} = \frac{\theta_{\mathrm{A}}(1-e^{i\Delta\phi})}{(1-i)\beta_2}\left(e^{\beta_2x} - e^{i\beta_2x}\right),
\end{aligned}
\end{equation}
where the exponents
\begin{gather}
\alpha_{2} = \left[2\pi\xi_{\perp}(1+\Gamma_0\Gamma)^{-1}/(B\tau)\right]^{1/4}e^{(7/8\pi) i}, \\
\beta_2 = \left[2\pi\xi_{\perp}(1+\Gamma_0\Gamma_{\mathrm{s}}-\Gamma_0\Gamma_{\mathrm{as}})^{-1}/(B\tau)\right]^{1/4}e^{(7/8\pi) i}.
\end{gather}
With $\zeta_{\mathrm{p}}$ and $\zeta_{\mathrm{m}}$ known, we can compute $\zeta_1$ and $\zeta_2$ using Eq.~(\ref{variable_change}). Substituting either $\zeta_1$ or $\zeta_2$ into Eq.~(\ref{U_analytic_zeta}), we obtain the swimming velocity at large $\eta$,
\begin{equation}\label{average_U_para_phaselag}
\begin{aligned}
\langle U \rangle &= \theta_{\mathrm{A}}^2\frac{L}{\tau}\frac{(2\pi)^{1/2} (2-\sqrt{2})}{8} \left(\frac{B \tau}{\xi_{\perp} L^4}\right)^{1/2} \left\{(1+\Gamma_0\Gamma)^{-1/2}(1+\cos\Delta\phi) + [1+\Gamma_0(\Gamma_{\mathrm{s}}-\Gamma_{\mathrm{as}})]^{-1/2}(1-\cos\Delta\phi) \right\} \\
&= \frac{1}{2}\langle U_0 \rangle \left\{(1+\Gamma_0\Gamma)^{-1/2}(1+\cos\Delta\phi) + [1+\Gamma_0(\Gamma_{\mathrm{s}}-\Gamma_{\mathrm{as}})]^{-1/2}(1-\cos\Delta\phi) \right\}.
\end{aligned}
\end{equation}
As shown in Fig.~(\ref{fig_S3}), Eq.~(\ref{average_U_para_phaselag}) fits well to the simulation result with $\Gamma_{\mathrm{s}}$ and $\Gamma_{\mathrm{as}}$ the fitting parameters.

\section{Effect of the drift flow}

\begin{figure}[t]
\centering
\includegraphics[bb= 0 0 340 150, scale=0.9, draft=false]{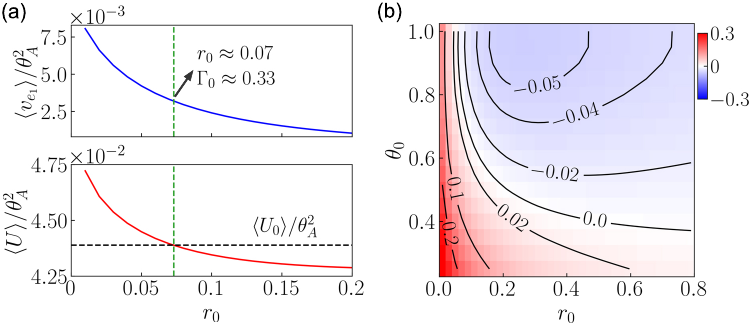}
\caption{Effect of the time-averaged drift flow $\langle v_{e_1}\rangle$ on the swimming performance. (a) Top panel: $\langle v_{e_1}\rangle$ as a function of $r_0$; bottom panel: $\langle U \rangle$ as a function of $r_0$. The dark dash line indicates the value of $\langle U_0 \rangle$ without iHIs. Other parameters are $\theta_0 = \theta_{\mathrm{A}} = 1.0$. The green dash line indicates the value of $r_0$ for which $\langle U \rangle = \langle U_0 \rangle $. (b) The relative change of efficiency due to iHIs, $(\mathcal{E}-\mathcal{E}_0)/\mathcal{E}_0$, as a function of $\theta_0$ and $r_0$ for $N=2$. }
\label{fig_S4}
\end{figure}

We illustrate the effect of the drift flow using the classic elastic scallop swimmer with $N = 2$. The two filaments are mirror symmetric about the central axis. First, we compare the time-averaged drift flow, $\langle v_{e_1}\rangle = \int_{0}^{\tau}v_{e_1}(t)\,dt/\tau$, with the time-averaged swimming speed $\langle U\rangle$ when varying the circle radius $r_0$. We set a large tilting angle $\theta_0 = 1.0$. As shown in Fig.~\ref{fig_S4}(a), iHIs enhance the swimming speed with $\langle U \rangle > \langle U_0\rangle$ due to the contribution of $\langle v_{e_1}\rangle$ for $r_0 \to 0$. As $r_0$ increases, both $\langle v_{e_1}\rangle$ and $\langle U\rangle$ decrease. When $r_0 \gtrsim 0.07$ ($\Gamma_0 \lesssim 0.33$), $\langle U \rangle$ becomes smaller than $\langle U_0 \rangle$. The impeded swimming is caused by the positive coupling strength $\Gamma$, as illustrated by the schematic shown in Fig. 2(c) inset of the main text. Figure~\ref{fig_S4}(b) shows the relative change of the efficiency, $(\mathcal{E}-\mathcal{E}_0)/\mathcal{E}_0$ as a function of $\theta_0$ and $r_0$. It is evident that for $r_0 \to 0$, $\mathcal{E} > \mathcal{E}_0$ for both small and large $\theta_0$. But for $r_0 \gtrsim 0.1$, $\mathcal{E} > \mathcal{E}_0$ only when $\theta_0$ is small. 
\begin{figure}[b]
\centering
\includegraphics[bb= 0 10 365 120, scale=0.8,draft=false]{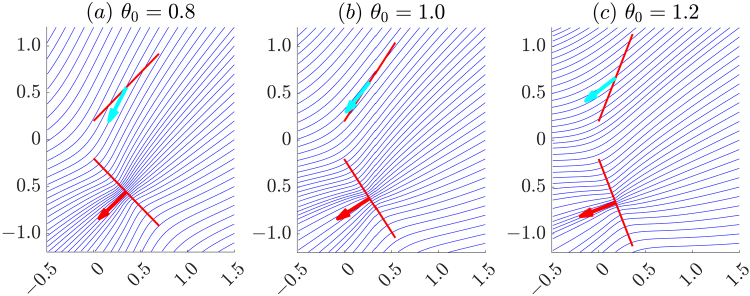}
\caption{Schematic of the coupling strength for $\theta_0 > 0$. The red arrows indicate point forces at the centers of the bottom filaments. The cyan arrows indicate the generated flows at the centers of the top filaments.}
\label{fig_S5}
\end{figure}

We also present in Fig.~\ref{fig_S5} a more detailed schematic of the coupling strength, corresponding to Fig. 2(c) inset of the main text, where the streamlines of a point force located at the filament center are shown. Note that the schematics, including the insets of Figs. 2(a) and 2(c) of the main text, only present a qualitative view. The calculation of the coupling strength requires one to integrate the flows generated by distributed force densities along the filaments.

\bibliographystyle{apsrev4-2}
\bibliography{reference}